\documentclass[aps,prb,longbibliography,floatfix,reprint]{revtex4-1}
\usepackage{amsmath}
\usepackage{graphicx}
\usepackage{amsfonts}
\usepackage{color}
\usepackage{placeins}

\begin{document}

\title{Theory of thermal expansion: Quasi-harmonic approximation and corrections from quasiparticle renormalization}

\author{ Philip B. Allen }
\email{philip.allen@stonybrook.edu}
\affiliation{ Department of Physics and Astronomy,
              Stony Brook University, 
              Stony Brook, New York 11794-3800, USA }

\date{\today}

\begin{abstract}
``Quasi-harmonic'' (QH) theory should not be considered a low-order 
theory of anharmonic effects in crystals, but should be recognized as an important
effect separate from ``true'' anharmonicity.  The original and 
widely used meaning of QH theory is to put $T=0$ volume-dependent 
harmonic phonon energies $\omega_Q(V)$ into the non-interacting phonon free energy.  This paper uses that meaning,
but extends it to include the use of $T=0$ $V$-dependent single-particle electron energies
$\epsilon_K(V)$.  It is demonstrated that the ``bare" quasiparticle (QP) energies $\omega_Q(V)$
and $\epsilon_K(V)$ correctly give the first-order term in
the $V$-dependence of the Helmholtz free energy $F(V,T)$.  Therefore, they give the leading order 
result for themal expansion $\alpha(T)$ and for the temperature-dependence of the bulk modulus $B(T)-B_0$.
However, neglected interactions which shift and broaden $\omega_Q$ with $T$, also shift
the free energy.  In metals, the low $T$ electron-phonon mass enhancement
of states near the Fermi level causes a shift in free energy that is similar in size to the electronic QH term.
Before $T$ reaches the Debye temperature $\Theta_D$, the mass renormalization essentially disappears,
and remaining electron-phonon shifts of free energy contribute only higher-order terms to thermal
expansion.  Similarly, anharmonic phonon-phonon interactions shift the free energy, but contribute to
thermal expansion only in higher order.  Explicit next order formulas are given for thermal expansion, which relate 
``true'' anharmonic and similar free energy corrections to quasiparticle self-energy shifts. 
The text below, except for Sec. IX, was already published in Modern Physics Letters B
Vol. 34, No. 2 (2020) 2050025.  Since then an important error was discovered.  The error is corrected
in Sec. IX, and will be published as an erratum in Modern Physics Letters B.
\end{abstract}

\maketitle


\section{Introduction}
\label{sec:intro}


Quasi-harmonic (QH) thermodynamics uses volume ($V$) dependence of ``bare'' quasiparticle (QP) energies
$\omega_Q(V)$ and $\epsilon_K(V)$ of phonons and electrons to compute the approximate free
energy $F(V,T)$.  It ignores the effect of interactions which introduce temperature ($T$)
dependence in measured QP energies $\omega_Q^{\rm QP}(V,T)$ and $\epsilon_K^{\rm QP}(V,T)$.
These interactions cause the free energy to be shifted from the QH approximation.  The question is,
how big is this shift compared to the influence of the QH shifts?   

Standard electronic structure theory
({\it e.g.} DFT \cite{Martin2016} and density functional perturbation theory DFPT \cite{Baroni2001,Baroni2010})
 give single-particle electron and phonon energies (the ``bare'' QP energies, which will also
be called ``QH energies'') $\epsilon_K(V)$ and $\omega_Q(V)$.
The symbol $K$ is short for $(\vec{k},n)$, the electron wavevector and band index.
The symbol $Q$ is short for $(\vec{q},j)$, the phonon wavevector and branch index.  The symbol
$V$ (and for more clarity, sometimes $\{V\}$) will mean the complete set of parameters 
$(\vec{a}, \ \vec{b},\ldots)$ needed to
define the structure, including any internal parameters.   Cubic crystals like rocksalt and zincblende
have only one structure parameter, volume.  A general crystal has $n$ parameters, all varying with temperature $T$.  
To compute $V(T)$ requires setting all external and internal strains $\{P\}=-\partial F(\{V\},T)/\partial \{V\}$ to zero.
 This determines the $T$-dependent structure $\{V\}(T)$.  The symbol $V_0$ or $\{V_0\}$ 
denotes the ``frozen lattice'' parameters, not quite the same as zero temperature parameters $\{V\}(T=0)$
which contain zero-point corrections.

 For phonons, the first QH correction, where $\omega_Q(V_0+\Delta V)-\omega_Q(V_0)$ is
approximated by $-\gamma_Q \omega_Q \Delta V/V$,
gives Gr\"uneisen \cite{Grueneisen1912,Grueneisen1926} theory of thermal expansion. 
The code GIBBS2 \cite{Otero2011} implements full quasiharmonic theory using density functional theory (DFT).
It is quite successful for equation-of-state and related thermodynamic calculations.  
QH calculations for MgO, Al, and C \cite{Otero-a-2011,Otero-b-2011}
fit experiment very well, especially when $V$ values are taken from experiment rather than DFT.
The methodology involves computing $V$-dependent phonon densities of states, without 
explicit evaluation of mode Gr\"uneisen parameters $\gamma_Q$, and going beyond low-order $V$-derivatives.

QH theory puts $\omega_Q(V)$ and $\epsilon_K(V)$ in the standard formulas for thermal free energy $F(V,T)$, 
of bare (or QH) phonons $F_{\rm  ph}^{\rm QH}$ and bare (or QH) electron single particle excitations $F_{\rm el}^{\rm QH}$.
It is unfortunately incorrect to insert the full renormalized energies into the noninteracting free energy $F_{\rm QH}$.
For electrons, the first order Taylor expansion of $\epsilon_K(V)$ in $\Delta V$ gives Mikura's theory \cite{Mikura1941} 
(rediscovered by Visvanathan \cite{Visvanathan1951}) of the electronic contribution to
thermal expansion.   

The terminology ``quasi-harmonic'' should be carefully distinguished from
``truly'' anharmonic.  In this paper, ``true anharmonicity'' is defined as the consequences of 
inter-atomic forces beyond harmonic, {\bf at fixed volume}.  
True anharmonicity shifts the phonon free energy $F(V,T)=F_{\rm QH}(V,T)+\Delta F_{\rm anh}(V,T)$.  
It therefore causes thermal expansion to deviate from the QH result contained in $F_{\rm QH}$.  This paper
examines corrections to QH theory from QP renormalization. 
The result is that QH theory does give correctly the leading volume dependence
of $F(V,T)$ needed for a correct lowest-order theory of thermal expansion.  How safe is it to ignore
the higher-order corrections from true anharmonicity?
A very careful study by Erba {\it et al.} \cite{Erba2015}
illuminates this issue.  For rocksalt-structure MgO, a weakly anharmonic material, the QH answer
is accurate to 1000K and starts to fail at higher $T$.  For isostructural CaO, with softer phonons and
therefore significant anharmonicity, breakdown of the QH prediction is seen already at 100K.
Silicon (see Ref. \onlinecite{Kim2018} and references therein) 
is another weakly anharmonic material where QH theory does well.
There is evidence that breakdown is very significant in NaBr \cite{Fultz2019}, because of strong
anharmonicity.  However, in PbTe which has ``giant'' anharmonicity \cite{Delaire2011},
QH theory by Skelton {\it et al.} \cite{Skelton2014} agrees over the measured \cite{Houston1968}
range $0<T<300$K.   ``True'' anharmonic corrections
are higher order in a parameter like $\epsilon=|\Delta\omega|/\omega$, where $\Delta\omega$ is the anharmonic
shift of the phonon with harmonic frequency $\omega$.  In most materials this is significantly smaller than 1,
but apparently not in some fairly simple ionic materials.  Even when $\epsilon$ is large, the
effect at $T<\Theta_D$ is not necessarily large.

Computations of first-order renormalization shifts of QP energies can be done with impressive reliability.  The ``true'' QP energies
differ from the lowest order ``bare'' QP energies in two ways:
\begin{equation}
\epsilon_K^{\rm QP}(V,T)=\epsilon_K(V_0)+\Delta\epsilon_K^{\rm QH}+\Delta\epsilon_K^{\rm QP}(V,T).
\label{eq:2sb}
\end{equation}
\begin{equation}
\omega_Q^{\rm QP}(V,T)=\omega_Q(V_0)+\Delta\omega_Q^{\rm QH}+\Delta\omega_Q^{\rm QP}(V,T).
\label{eq:2sa}
\end{equation}
The QH shift is simply the bare QP energy shifted from the $T=0$ frozen lattice parameters $V_0$ to
the actual volume $V=V(T)$,
\begin{equation}
\Delta\epsilon_K^{\rm QH}=\epsilon_K(V)-\epsilon_K(V_0); \ \ 
\Delta\omega_Q^{\rm QH}=\omega_Q(V)-\omega_Q(V_0).
\label{eq:}
\end{equation}
The QP shift is the $T$-dependent renormalization, obtained typically from an electron or phonon self-energy
calculation, done at a fixed volume $V$.  In priciple, the self-energy shift, $\Delta\omega_Q^{\rm QP}(V,T)$
or $\Delta\epsilon_K^{\rm QP}(V,T)$,
should be recomputed at each volume of interest, namely the $V$ of the system at temperature $T$.

The outline of the rest of the paper is: Sec. \ref{sec:conf} describes confusions that will be addressed.  Sec. \ref{sec:oom}
is about small parameters and orders of magnitude.  Sec. \ref{sec:QH} derives QH formulas
for thermal expansion.  Sec. \ref{sec:QPth} gives approximate thermodynamic formulas that include
renormalizations found in QP theory.  Sec. \ref{sec:ren} derives approximate corrections to thermal
expansion from QP renormalization.  Sec. \ref{sec:conc} summarizes the main conclusions.  An appendix
gives mathematics useful for relating free energy formulas to entropy.  
The new things in this paper are, first, the electron QP mass-enhancement correction to low $T$
metal thermal expansion has not been mentioned before; second,
explicit formulas correcting thermal expansion for QP renormalization are new.

\section{Confusions and Resolutions}
\label{sec:conf}

The first confusion is incomplete agreement about the definition of QH.  
Gr\"uneisen \cite{Grueneisen1912,Grueneisen1926}
used $V$-dependence of $\omega_Q(V)$ to obtain thermal expansion, $\alpha(T)$. 
This was designated as QH theory by Barron \cite{Barron1961} and
Leibfried and Ludwig \cite{Leibfried1961} (BLL) in 1961.
Unfortunately, Cowley \cite{Cowley1966I,Cowley1968} in 1966 gave a different definition of QH,
namely using QP energies $\omega_Q^{\rm QP}(V,T)$ in the theory of anharmonicity.  
The original BLL meaning is widely adopted, but the Cowley
 meaning still appears occasionally.  Sometimes the QH method is used without being
 given a name.  For example, Wallace \cite{Wallace1972} gives much of the theory and
 never uses the word quasiharmonic.
 
 A second confusion is that QH results are sometimes confused
with a first approximation to a full anharmonic theory, and criticized for not dealing
correctly with ``true'' anharmonicity.  This is mistake has 
 undermined appreciation of the true validity of QH theory as the leading approximation for $V$-dependence of $F(V,T)$.
 
 A third confusion arises because the QH shifts ($\Delta\epsilon_K^{\rm QH}$ and $\Delta\omega_Q^{\rm QH}$)
are similar in size to the QP shifts ($\Delta\epsilon_K^{\rm QP}$ and $\Delta\omega_Q^{\rm QP}$).
This raises the question of why QH theory should work for anything.

 A fourth point is that the QH electronic
 contribution to thermal structure shifts is often omitted.  This contribution, introduced in 1941
 by Mikura \cite{Mikura1941}, is only important in metals, and then primarily at low $T$.
 The correct inclusion of $V$-dependence of electronic band energies clarifies the fact that
 QH theory has very little to do with true anharmonicity.
 
This paper aims to answer these issues and to  reaffirm the validity of QH
theory. 

 \section{Orders of magnitude}
 \label{sec:oom}
 
 There are two energy scales, labeled $\hbar\omega_{\rm ph}$ and $E_{\rm el}$, which differ by
 typically two orders of magnitude ({\it i.e.} $\hbar\omega_{\rm ph}/E_{\rm el} \sim \sqrt(m/M)\sim 10^{-2}$,
 where $m$ and $M$ are electron and ion mass).  
 What are the approximate energy scales of the shifts
 $\Delta\epsilon_K^{\rm QP}$ and $\Delta\omega_Q^{\rm QP}$?  
 
 First consider the anharmonic renormalization
 of the phonon energy, $\Delta\omega_{\rm ph}^{\rm anh}\approx |V_3|^2/ \hbar\omega_{\rm ph}$.  
 How does this compare with
 $\omega_{\rm ph}$?  The anharmonic coupling is $V_3 \sim (d^3 U/du^3)u^3$.  The third derivative is
 $d^3 U/du^3 \sim E_{\rm el}/a^3$, where $a$ is a lattice constant.  The average lattice displacement is
 governed by $<u^2>\sim \hbar/2M\omega_{\rm ph}$.  Putting these together, we get
\begin{equation}
\frac{\Delta \omega_{\rm ph}^{\rm anh}}{\omega_{\rm ph}} \sim \frac{\hbar\omega_{\rm ph}}{E_{\rm el}}\sim 0.01
\label{eq:dEC}
\end{equation}
This is an underestimate, for at least two reasons.  First, there are inevitable complications, such as three 
phonon branches, which could enhance the estimate by as much as $3^2$, and second, at higher
$T$, the displacement $<u^2>$ increases linearly with $k_B T/\hbar\omega_{\rm ph}=T/\Theta_D$.  For ordinary materials,
a safer estimate is $0.02T/\Theta_D<\Delta \omega_{\rm ph}^{\rm anh}/\omega_{\rm ph}<0.1T/\Theta_D$.

Now consider the Coulomb renormalization of the electron energy, $\Delta E_{\rm el}^{\rm Coul}\sim |v_C|^2/E_{\rm el}$.
The Coulomb matrix element $|v_C|$ is of order $E_{\rm el}$, so the fractional shift is of order 1.
But it is the job of DFT to include all relevant $T=0$ Coulomb renormalization in the bands $\epsilon_K$.
Only the thermal alteration of Coulomb renormalization is left to include in $\Delta\epsilon_K^{\rm QP}(V,T)$,
and this is usually unnecessary.  The lifetime broadening caused by Coulomb scattering is
smaller by $(\hbar/\tau_C)/E_{\rm el} \sim (k_B T/E_{\rm el})^2$.  A similar, and negligible thermal
Coulomb shift of the QP energy can be expected.

Finally, consider electron-phonon interactions.  For both electron and phonon renormalization, the
energy denominator in second order perturbation theory is $\sim E_{\rm el}$.  The numerator is
$|V_{\rm ep}|^2 \sim (dU/du)^2 <u^2> \sim E_{\rm el}\hbar\omega_{\rm ph}$.  Therefore the result is
\begin{equation}
\Delta_{\rm ep} \epsilon_K^{\rm QP}\sim \Delta_{\rm ep} \hbar\omega_Q^{\rm QP} \sim \hbar\omega_{\rm ph}.
\label{eq:dEep}
\end{equation}
For electronic states $\epsilon_K$, this is a small effect.  It is omitted in band structure calculations, but appears in
experiment as zero-point and thermal shifts of band gaps \cite{Allen1976,Ponce2015}.
These can be measured and computed, and make a correction
of order $\hbar\omega_{\rm ph}$ or a few $k_B T$.  However, in
metals at low $T$ ({\it i.e.} when thermal excitation of electrons $E_{\rm el}^{\rm th}\sim k_B T$ is significantly
smaller than $\hbar\omega_{\rm ph}$), there is an important, additional, non-adiabatic effect caused by the
singularity at the sharp Fermi edge $\epsilon_K\approx E_F$.  This causes a ``mass renormalization'' 
$\epsilon_K \rightarrow \epsilon_K/(1+\lambda_K(T))$.
The parameter $\lambda\equiv<\lambda_K(T=0)>$ is the superconducting coupling constant, usually
in the range $0.1 < \lambda < 2.0$.  As $T$ increases, $\lambda_K(T)$ decreases rapidly; it becomes
negligible before $T=\Theta_D$. \cite{Grimvall1981}
This mass renormalization is actually a shift of bands of the expected order ($\sim\hbar\omega_{\rm ph}$), but only
for electron states within $\sim\hbar\omega_{\rm ph}$ of $E_F$.  This enhances the electron
density of states at $T=0$ (at the Fermi level), which in turn causes 
the familiar $(1+\lambda)$ enhancement of the low-$T$ electronic specific heat.  
It should also appear in low-$T$ thermal expansion of metals.  It is the only 
common case where QH theory gets an order 1 correction from QP renormalization.

For phonons, however, the electron-phonon renormalization ($\sim \omega_{\rm ph}$)
is a shift of order 1.  It is the job of DFT (more specifically, density-functional perturbation theory, DFPT \cite{Baroni2001}) 
to include $T=0$ electron-phonon renormalization in the calculation of ``bare'' phonon bands.  Additional thermal
renormalization also occurs, with both lifetime broadening and phonon energy shifts.  The lifetime
broadening of phonons in metals is $\sim (\hbar\omega_{\rm ph}/E_{\rm el})k_B T$, similar in size to (but usually smaller than)
anharmonic lifetime broadening.  Therefore the residual thermal shift $\Delta\omega_Q^{\rm QP}(T)$ of phonon QP energy
from electron-phonon interactions, omitted in phonon
band calculations, is similar in magnitude to anharmonic shifts.  Both effects give higher-order corrections
to thermal expansion.

\section{Quasi-harmonic Thermal Expansion}
\label{sec:QH}

Some thermodynamic relations are
\begin{equation}
P(V,T)=-\left(\frac{\partial F}{\partial V}\right)_T; \ \ V(P,T)=\left(\frac{\partial G}{\partial P}\right)_T,
\label{eq:eos}
\end{equation}
where $F$ and $G$ are the Helmholz and Gibbs free energies.
These are versions of the equation of state.  Further relations are
\begin{equation}
\alpha = \frac{1}{V}\left(\frac{\partial V}{\partial T}\right)_P = \frac{1}{V}\left(\frac{\partial^2 G}{\partial T\partial P}\right)_{PT},
\label{eq:alph}
\end{equation}
\begin{equation}
B_T = -V\left(\frac{\partial P}{\partial V}\right)_T=V\left(\frac{\partial^2 F}{\partial V^2}\right)_T.
\label{eq:BT}
\end{equation}
where $\alpha$ is the coefficient of volume thermal expansion and $B_T$ is the isothermal bulk modulus.
A curious connection is the relation
\begin{equation}
\alpha B_T = -\left(\frac{\partial^2 F}{\partial V \partial T}\right)_{TV},
\label{eq:Fultz}
\end{equation}
which is given as Eq. 1.35 of ref. \onlinecite{Wallace1972}.

The QH energies generate QH free energies,
\begin{eqnarray}
F_{\rm QH} &=& -k_B T \ln (Z_{\rm el}^{\rm bare} \times Z_{\rm ph}^{\rm bare}) \nonumber \\
&=& F_{\rm el}^{\rm QH}+F_{\rm ph}^{\rm QH}.
\label{eq:Fbare}
\end{eqnarray}
Technically, the electron part is not $F(V,T,N)$ but the ``grand potential'' $\Omega(V,T,\mu)$.  For the phonon part,
there is no difference ($N_{\rm ph}$ is not conserved, so there is no $\mu_{\rm ph}$), 
and for the electron part, in this paper the difference is irrelevant.  The symbol $F$ will
be used instead of $\Omega$.  The electron part is
\begin{eqnarray}
&&F_{\rm el}^{\rm QH}(V,T)=k_B T \sum_K \ln(1-f_{K0}(V,T)) \nonumber \\
&=&\sum_K^{\rm occ} (\epsilon_K(V)-\mu(V))+k_B T\sum_K \ln\frac{1-f_{K0}(V,T)}{1-f_{K0}(V,0)}\nonumber \\
&\rightarrow&E_{\rm el}(V) +k_B T\sum_K \ln\frac{1-f_{K0}(V,T)}{1-f_{K0}(V,0)} \nonumber \\
&& \ \ \equiv E_{\rm el}(V)+\Delta F_{\rm el}^{\rm QH}(V,T).
\label{eq:Fbareel}
\end{eqnarray}
Here $f_K$ is the Fermi-Dirac distribution.  The additional
subscript 0 indicates that the energy $\epsilon_K(V)$
appearing in $f_{K0}$ is the QH energy, independent of $T$.  The phonon part is
\begin{equation}
F_{\rm ph}^{\rm QH}(V,T)=\sum_Q \frac{\hbar\omega_Q(V)}{2}-k_B T \sum_Q \ln(1+n_{Q0}(V,T)).
\label{eq:Fbareph}
\end{equation}
Here $n_Q$ is the Bose-Einstein distribution.  The 
subscript 0 indicates that the energy $\omega_Q(V)$
appearing in $n_{Q0}$ is the QH energy, independent of $T$.

The first term in Eq. \ref{eq:Fbareph} is the zero-point phonon energy.  In the second line of
Eq. \ref{eq:Fbareel}, the zero temperature part has been added and subtracted in order to
make the zero-point electronic contribution explicit.  In the third and fourth lines of Eq. \ref{eq:Fbareel},
the zero-point electronic contribution has been replaced by a more accurate electronic
$T=0$ frozen lattice energy, $ E_{\rm el}(V)$.  This is computed, for example, by DFT.
It contains additional energy, beyond the single-particle zero-point part of the first two lines of Eq. \ref{eq:Fbareel}.  

Now Taylor-expand in powers of the lattice displacements around the frozen lattice structure $V_0$.
The zeroth order term, $F_{\rm QH}(V_0,T)$, contains the large $T$-independent piece $E_{\rm el}(V_0)$,
plus thermal shifts  $\Delta F_{\rm el}^{\rm QH}(V_0,T)+\Delta F_{\rm ph}^{\rm QH}(V_0,T)$.  
This $(V-V_0)$-independent term does not affect thermal expansion.  The first contributing terms in the expansion are
\begin{eqnarray}
&&F_{\rm QH}(V,T)-F_{\rm QH}(V_0,T)=\frac{B_0 V_0}{2}\left( \frac{V-V_0}{V_0}\right)^2 \nonumber \\
&& \ \ +V_0\left[\frac{\partial\Delta F_{\rm el}^{\rm QH}}{\partial V}+\frac{\partial F_{\rm ph}^{\rm QH}}{\partial V}\right]_{V_0}
\left( \frac{V-V_0}{V_0}\right)+\ldots.
\label{eq:lowexp}
\end{eqnarray}
By definition, the large term $E_{\rm el}(V)$ is minimum at $V=V_0$, so does not contribute
in first order in $(V-V_0)$.  Its second derivative gives $B_0$,
\begin{equation}
B_0 = V_0\left( \frac{\partial^2 E_{\rm el}}{\partial V^2}\right)_{V^\ast}.
\label{eq:B}
\end{equation}
Here $B_0$ is the leading-order strain tensor, equal to a scalar, the bulk modulus, in simple crystals.
It is uncertain what volume $V^\ast$ should be used when the derivatives are taken.  The convenient choice
$V^\ast = V_0$ will be used here.  The corrections from different $^\ast$ choices are small.  
The second derivatives of $\Delta F_{\rm el}^{\rm QH}(V,T)+\Delta F_{\rm ph}^{\rm QH}(V,T)$
are omitted because they are small compared to $B_0 V$.
Stopping the expansion here gives the standard Gr\"uneisen theory.  Recent advances in computation have allowed 
reliable evaluation of higher-order effects, beyond Gr\"uneisen.  
These give noticable corrections at higher $T$, as will be mentioned below.

The quasiharmonic equation of state is given by $P(V,T)=-(\partial F_{\rm QH}/\partial V)_T$).
The zero pressure $P=0$ structure parameters minimize the free energy 
$(\partial F/\partial V)_T  = 0$).  This gives the first-order QH result,
\begin{equation}
\frac{V_1^{\rm QH}(T)-V_0}{V_0}=-\frac{1}{B_0}\left[\left(\frac{\partial\Delta F_{\rm el}^{\rm QH}}{\partial V}\right)_T
+\left(\frac{\partial F_{\rm ph}^{\rm QH}}{\partial V}\right)_T\right]_{V_0},
\label{eq:DV1}
\end{equation}
For the $P>0$ result, add $P$ to the terms inside $[ \ ]$.  That gives the first-order QH equation of state.
Now introduce electron and phonon Gr\"uneisen parameters.  They are in general vectors,  
and $1/B_0$ is a tensor.  Gr\"uneisen \cite{Grueneisen1926} already recognized the 
vector nature of his parameters:
\begin{equation}
\gamma_K\equiv -\left(\frac{V}{\epsilon_K-\mu}\frac{d(\epsilon_K-\mu)}{dV}\right)_{V_0}
\label{eq:gK}
\end{equation}
\begin{equation}
\gamma_Q\equiv -\left(\frac{V}{\omega_Q}\frac{d\omega_Q}{dV}\right)_{V_0}
\label{eq:gQ}
\end{equation}
These are computed from the bare QP ({\it i.e.} the QH) energies.
Then the two parts of the QH thermal expansions are
\begin{equation}
\left( \frac{V_1-V_0}{V_0}\right)_{\rm el}^{\rm QH}=\frac{1}{B_0 V_0}\sum_K (\epsilon_K-\mu)\gamma_K (f_{K0}(V,T)-f_{K0}(V,0)) 
\label{eq:QHel}
\end{equation}
\begin{equation}
\left( \frac{V_1-V_0}{V_0}\right)_{\rm ph}^{\rm QH}=\frac{1}{B_0 V_0} \sum_Q\hbar\omega_Q\gamma_Q (n_Q+1/2).
\label{eq:QHph}
\end{equation}
This is Gr\"uneisen limit of QH theory \cite{Grimvall1986}.
Equation \ref{eq:QHel} is the Mikura theory \cite{Mikura1941} and Eq. \ref{eq:QHph} is the Gr\"uneisen
theory \cite{Grueneisen1912,Grueneisen1926}.  The electronic part
is usually omitted.  It dominates in metals at low $T$, but is otherwise much smaller than the
vibrational part, especially in non-metals.   
The full expression for $\Delta V$ (which includes zero-point effects) is not usually given in the literature;
Refs. \onlinecite{Stern1958,Ziman1960,Leibfried1961,Pavone1994,Huang2016,Wallace2002} give parts or all
of the answer.  More often the $T$ derivative $\alpha(T)=(dV/dT)/V$ is given,
\begin{equation}
\alpha_{\rm QH}(T)=\frac{1}{B_0} \left[ \sum_K C_K^{\rm el}(T) \gamma_K +\sum_Q C_Q^{\rm ph}(T) \gamma_Q \right],
\label{eq:aQHph}
\end{equation}
where $V_0 C_K^{\rm el}=(\epsilon_K-\mu)\partial f_K/\partial T$ is the mode $K$ contribution to the electron
specific heat, and $V_0 C_Q^{\rm ph}=\hbar\omega_Q \partial n_Q/\partial T$ is the mode $Q$ contribution to the phonon
specific heat.  The corresponding formula for the first $T$-dependent correction to the bulk modulus is
\begin{eqnarray}
B_{\rm QH}(T)&-& B_0 = \sum_Q \left[U_Q(T) \gamma_Q^\prime -T C_Q \gamma_Q^2  \right] \nonumber \\
&+& \sum_K \left[[U_K(T)-U_K(0)] \gamma_K^\prime -T C_K \gamma_K^2 \right]
\label{eq:BQH}
\end{eqnarray}
\begin{equation}
\gamma_Q^\prime \equiv \left( \frac{V^2}{\omega_Q} \frac{d^2 \omega_Q}{dV^2} \right)_{V_0} ; \ \ 
\gamma_K^\prime \equiv \left( \frac{V^2}{(\epsilon_K-\mu)} \frac{d^2 (\epsilon_K-\mu)}{dV^2} \right)_{V_0}
\label{eq:gphp}
\end{equation}
where $V_0 U_Q \equiv \hbar\omega_Q (n_Q + \tfrac{1}{2})$ is the mode $Q$ contribution to the volumetric
thermal vibrational  energy, and $V_0 U_K \equiv \epsilon_K f_K$ is the mode $K$ contribution
fo the non-interacting volumetric thermal electron energy.  I did not find these leading order formulas
for $\Delta B$ in previous literature.
Wu and Wentzcovitch \cite{Wu2011} give general formulas for the tensor elastic constants in QH theory and
classical approximation.  Davies \cite{Davies1974} and others have done it
with quantized lattice vibrations.

At $T>\Theta_D$, the phonon part of Eq. \ref{eq:aQHph} (Gr\"uneisen theory) 
gives a thermal expansion $\alpha(T)$ that saturates at 
 $(k_B/B_0 V_0)\sum_Q \gamma_Q$, similar to the behavior of the harmonic specific heat $C(T)$.  
 Experimental $C(T)$ usually saturates fairly cleanly.  However, $\alpha(T>\Theta_D)$ is  
 often less constant than $C(T>\Theta_D)$.  This does not necessarily indicate that
 true anharmonic corrections have begun entering.  The lowest order version, Eq. \ref{eq:lowexp},
 of QH theory needs to be extended to higher powers of $\Delta V$.  
 Rather than further terms in Taylor series, it is more common simply to do the full calculation
 in Eqs. \ref{eq:Fbareel} and \ref{eq:Fbareph} numerically.  At each chosen volume, the QH free energy
 is calculated {\it versus} $T$.  The value of $T$ that minimizes $F_{\rm QH}$ gives  
 a discreet point on the QH $V(T)$ curve.  It was done, for example, with the help of model
 equations of state, by Kantorovich \cite{Kantorovich1995-II} in 1995 for KCl and NaCl.  Advances in computation now
  allow reliable model-free calculations of this kind; for example, the work by
  Skelton {\it et al.} \cite{Skelton2014} for PbTe and Erba {\it et al.} \cite{Erba2015} for MgO and CaO.
  Deviations from constant $\alpha(T>\Theta_D)$ are often quite well described.  The full QH theory shows that
  anharmonic corrections may be unnecessary for a weakly anharmonic material like MgO until
  $T$ is significantly higher than $\Theta_D$.  The only advantage of using the Gr\"uneisen expansion is
  that Eqs. \ref{eq:QHel} and \ref{eq:QHph} have an appealing simplicity that disappears in the more
  complete version of QH theory.

\section{Quasiparticle Thermodynamics}
\label{sec:QPth}

The aim is to compute renormalization corrections (anharmonic and electron-phonon) beyond QH
to the $V$-dependence of $F(V,T)$.  
This problem has been addressed by many authors, for example, Werthamer
\cite{Werthamer1970} and G\"otze and Michel \cite{Gotze1968}.
The straightforward route is to use
Feynman-Dyson-Matsubara perturbation theory for such corrections. 
For anharmonic phonon interactions, the answer
derived by Liebfried and Ludwig \cite{Leibfried1961} can be used for this purpose.
Unfortunately, the corresponding correction to $F(V,T)$ from
electron-phonon interactions double-counts the electron-phonon renormalization effects on phonons, which
were already included in $F_{\rm QH}^{\rm ph}$.  Therefore an alternate route is
needed.  One can find the desired corrections to entropy $S$, and then integrate to get $F$.

For gases of independent quasiparticles obeying Fermi-Dirac or Bose-Einstein statistics,
entropy can be evaluated by counting the number of ways of distributing their occupation.
The formulas are \cite{Lifshitz1980}
\begin{equation}
S_{\rm el}=-k_B\sum_K [f_K \ln f_K +(1-f_K)\ln(1-f_K)].
\label{eq:Sel}
\end{equation}
\begin{equation}
S_{\rm ph}=k_B \sum_Q [(n_Q+1)\ln(n_Q+1)-n_Q\ln n_Q],
\label{eq:Sph}
\end{equation}
Although rigorously true only for non-interacting particles, there is justification for 
using these formulas for gases with reasonably well-defined quasiparticles.  The distributions
$f_K$ and $n_Q$ then have $T$-dependent energies
\cite{Barron1963,Wallace1972}.   The renormalization corrections to $S$ are given correctly
to lowest order, and effects of higher order in perturbation theory are partially captured.
Then the $T$-dependent part of the free energy can be obtained from
\begin{equation}
F(V,T)-F(V,0)=-\int_0^T dT^\prime S(V,T^\prime).
\label{eq:FSint}
\end{equation}
If QH energies (independent of $T$) are used in the distributions $f_K$ and $n_Q$ in Eqs. \ref{eq:Sel} and \ref{eq:Sph},
the $dT^\prime$ integrals can be done explicitly, and give Eqs. \ref{eq:Fbareel} and \ref{eq:Fbareph}, except
that the zero-point contributions have been subtracted off.  In other words, they are exactly as
expected from Eq. \ref{eq:FSint}.  As a reality check, a method for doing the integration is explained in the appendix.

\section{Corrections to QH theory from QP renormalization}
\label{sec:ren}

QH theory puts 
the first two terms of the energies in Eqs. \ref{eq:2sb} and \ref{eq:2sa}
into the free energy Eqs. \ref{eq:Fbareel} and \ref{eq:Fbareph}.
To get the renormalization corrections $\Delta F_{\rm renorm}$
from $\Delta\epsilon_K^{\rm QP}(V,T)$ and $\Delta\omega_Q^{\rm QP}(V,T)$,
use the entropy formulas \ref{eq:Sel} and \ref{eq:Sph} with the full QP energies in Eqs.  \ref{eq:2sb} and \ref{eq:2sa}.
Then Taylor expand to first order in $\Delta\epsilon_K^{\rm QP}(V,T)$ and $\Delta\omega_Q^{\rm QP}(V,T)$.
The zeroth order terms can be integrated over $T^\prime$ as in Eq. \ref{eq:FSint} to reproduce the QH results.
The first order terms can then be numerically integrated over $T^\prime$ to give the corrections
 $\Delta F_{\rm renorm}(V,T)$.  Then the volume derivative gives the correction to $(V-V_0)/V_0$.
 
 To be specific, here is the electronic contribution from QP renormalization:
\begin{eqnarray}
\frac{\partial \Delta F_{\rm el}^{\rm QP}}{\partial V}&=& \int_0^T dT^\prime \sum_K \frac
{\partial[(1-f_K)\ln(1-f_K) + f_K\ln f_K]}{\partial(\epsilon_K -\mu)} \nonumber \\
&& \ \ \ \ \ \times k_B \frac{\partial \Delta(\epsilon_K -\mu)_{\rm QP}}{\partial V}.
\label{eq:dFel}
\end{eqnarray}
The shift $\Delta(\epsilon_K -\mu)_{\rm QP}$ is the renormalization $\Delta\epsilon_K^{\rm QP}(V,T^\prime)$ in Eq. \ref{eq:2sb},
minus the shift $\Delta\mu$ of the chemical potential.  A Gr\"uneisen-like parameter
$\Delta\gamma_K$ can be defined,
\begin{equation} 
\Delta\gamma_K = -\left( \frac{V}{(\epsilon_K-\mu)}\frac{\partial \Delta(\epsilon_K -\mu)_{\rm QP}}{\partial V}\right)_{V^\ast=V_0}.
\label{eq:Dg}
\end{equation}
This is smaller than the analogous Eq. \ref{eq:gK} for the ordinary electronic Gr\"uneisen parameter.
The fractional volume shift $-V[\partial \Delta(\epsilon_K -\mu)_{\rm QP}/\partial V]/(\epsilon_K -\mu)_{\rm QP}$
 of the quasiparticle renormalization is probably similar in size to the ordinary Gr\"uneisen parameter,
 the fractional volume shift $-V[\partial \Delta(\epsilon_K -\mu)_0/\partial V]/(\epsilon_K -\mu)_0$.
 But since both are normalized to $(\epsilon_K -\mu)_0$, $\Delta\gamma_K$ is smaller than $\gamma_K$.
 Once again, there is uncertainty about the
 volume $V^\ast$ to be used, but the convenient value $V_0$ is chosen here.
 Then Eq. \ref{eq:dFel} can be written
\begin{equation}
\frac{\partial \Delta F_{\rm el}^{\rm QP}}{\partial V}=-\int_0^T dT^\prime \sum_K C_K(T^\prime)
\Delta\gamma_K(T^\prime),
\label{eq:DelQP}
\end{equation}
This can now be compared to the corresponding QH formula (compare Eqns. \ref{eq:DV1} and \ref{eq:QHel}),
\begin{equation}
\frac{\partial  F_{\rm el}^{\rm QH}}{\partial V}=
-\frac{1}{V_0}\sum_K (\epsilon_K-\mu)\gamma_K (f_{K0}(V,T)-f_{K0}(V,0))
\label{eq:DelQH}
\end{equation}
%
These formulas are very similar.  If Gr\"uneisen-type parameters are ignored, then Eq. \ref{eq:DelQH}
is the thermal electron energy per volume, while Eq. \ref{eq:DelQP} is the $T$-integrated electronic
specific heat (the same thing, except for slightly different volume normalizations).  But the
Gr\"uneisen-type parameters are quite different.  The ratio $\Delta\gamma_K /\gamma_K$ is
expected to be similar in size to $\Delta (\epsilon_K -\mu)_{\rm QP}/(\epsilon_K -\mu)$, a small number.
Therefore, the QP correction to the QH $dF/dV$
is small, and (except for the low-$T$ mass renormalization) QH theory is justified for the electron part. 

 The phonon contribution from QP renormalization is
\begin{eqnarray}
\frac{\partial \Delta F_{\rm ph}^{\rm QP}}{\partial V}&=& -\int_0^T dT^\prime \sum_Q \frac
{\partial[(n_Q+1)\ln(n_Q+1) + n_Q\ln n_Q]}{\partial(\omega_Q)} \nonumber \\
&& \ \ \ \ \ \times k_B \frac{\partial \Delta\omega_Q^{\rm QP}}{\partial V}.
\label{eq:dFph}
\end{eqnarray}
The shift $\Delta\omega_Q^{\rm QP}$ is the renormalization $\Delta\omega_Q^{\rm QP}(V,T^\prime)$ in Eq. \ref{eq:2sa}.
Define a Gr\"uneisen-like parameter $\Delta\gamma_Q$,
\begin{equation}
\Delta\gamma_Q = -\left( \frac{V}{\omega_Q}\frac{\partial \Delta\omega_Q^{\rm QP}}{\partial V}\right)_{V^\ast=V_0}
\label{eq:Dgph}
\end{equation}
This is smaller than the analogous Eq. \ref{eq:gQ} for the ordinary phonon Gr\"uneisen parameter.
The fractional volume shift $-V(\partial\Delta\omega_Q^{\rm QP}/\partial V)/\Delta\omega_Q^{\rm QP}$
should be similar in size to $\gamma_Q$.  But $\Delta\gamma_Q$ is normallized to the larger
harmonic frequency $\omega_Q$.
There is another important difference between 
this Gr\"uneisen-like parameter and the ordinary Gr\"uneisen parameter of Eq. \ref{eq:gQ}: $\Delta\gamma_Q$
is $T$-dependent (increases with $T$ as anharmonicity increases), while $\gamma_Q$ is a constant, independent
of $T$.  The same may be true of the electronic versions.

Then Eq. \ref{eq:dFph} becomes
\begin{equation}
\frac{\partial \Delta F_{\rm ph}^{\rm QP}}{\partial V}=-\int_0^T dT^\prime \sum_Q C_Q(T^\prime)
\Delta\gamma_Q(T^\prime),
\label{eq:DphQP}
\end{equation}
This can now be compared to the corresponding QH formula (compare Eqns. \ref{eq:DV1} and \ref{eq:QHph}),
\begin{equation}
\frac{\partial F_{\rm ph}^{\rm QH}}{\partial V}=-\frac{1}{V_0} \sum_Q\hbar\omega_Q\gamma_Q (n_Q+1/2)
\label{eq:DphQH}
\end{equation}
These two equations are parallel to Eqs. \ref{eq:DelQP} and \ref{eq:DelQH}, except that Eq. \ref{eq:DphQH} has an
extra zero-point energy contribution (the 1/2).  This does not alter the observation that Eqs. \ref{eq:DphQP}
and \ref{eq:DphQH} are very similar in magnitude except that $\Delta\gamma_Q/\gamma_Q$ has the magnitude
of $\Delta\omega_Q^{\rm QP}/\omega_Q$, based on similar sensitivity to volume change.  This demonstrates
that QP corrections to the phonon QH volume derivative of $F$ are small. 

Similar results can be found for the QP corrections to the
QH thermal shift of $B=V\partial^2 F/\partial V^2$. Except for the low $T$ electronic mass renormalization correction,
the QP corrections to $B$ are normally smaller than the QH corrections given in Eq. \ref{eq:BQH}. 

The explicit quasiparticle correction formulas \ref{eq:DelQP} and \ref{eq:DphQP} 
have not been published before.  They have potential use for thinking about
or computing higher order thermal expansion or $B(T)$ corrections.  Consider the next order
corrections to the expansion of Eq. \ref{eq:lowexp},
\begin{eqnarray}
F(V,T)&-&F(V_0,T)=\frac{B_0 V_0}{2}\left( \frac{V-V_0}{V_0}\right)^2 +\frac{B_0^\prime V_0}{6}
\left( \frac{V-V_0}{V_0}\right)^3 \nonumber \\
&+&V_0\left[\frac{\partial\Delta F_{\rm el}^{\rm QH}}{\partial V}+\frac{\partial F_{\rm ph}^{\rm QH}}{\partial V}\right]_{V_0}
\left( \frac{V-V_0}{V_0}\right) \nonumber \\
&+& \frac{V_0^2}{2}\left[\frac{\partial^2 \Delta F_{\rm el}^{\rm QH}}{\partial V^2}
+\frac{\partial^2 F_{\rm ph}^{\rm QH}}{\partial V^2}\right]_{V_0}
\left( \frac{V-V_0}{V_0}\right)^2 \nonumber \\
&+&V_0\left[\frac{\partial\Delta F_{\rm el}^{\rm QP}}{\partial V}+\frac{\partial \Delta F_{\rm ph}^{\rm QP}}{\partial V}\right]_{V_0}
\left( \frac{V-V_0}{V_0}\right) +\ldots.
\label{eq:medexp}
\end{eqnarray}
Three new terms have been added.  (1) The frozen lattice electron energy $E_{\rm el}(V)$ now has a third derivative
term $B_0^\prime=V_0(\partial B_0/\partial V)_{V_0}$.  (2) The QH electron energy $\epsilon_K(V)$ and QH
phonon energy $\omega_Q(V)$ volume dependence is now expanded to second order.
(3) The quasiparticle corrections of Eqs. \ref{eq:DelQP} and \ref{eq:DphQP} are now kept in first order.  
The volume that minimizes this expression (or, for $P>0$, that satisfied $P=-(\partial F/\partial V)_T$) 
contains the next order corrections, both QH and QP, to the  
first-order (Gr\"uneisen) version of QH theory.

Suppose the first two corrections are smaller than the third, and temporarily omitted.  Then the minimum
free energy is at
\begin{equation}
\left( \frac{V-V_0}{V_0}\right)(T)=-\frac{1}{V_0}\left[ \frac{\partial \Delta F^{\rm QH}}{\partial V} 
+ \frac{\partial \Delta F^{\rm QP}}{\partial V} \right].
\label{eq:DFtot}
\end{equation}
The corresponding formula for thermal expansion is
\begin{equation}
\alpha(T) \approx \alpha_{\rm QH} + \Delta \alpha_{\rm QP}
\label{eq:a}
\end{equation}
where $ \Delta \alpha_{\rm QP}=\Delta \alpha_{\rm QP}^{\rm el}+\Delta \alpha_{\rm QP}^{\rm ph}$, and
\begin{equation}
\Delta \alpha_{\rm QP}(T)=\frac{1}{B_0} \left[ \sum_K C_K^{\rm el}(T) \Delta \gamma_K 
+\sum_Q C_Q^{\rm ph}(T) \Delta \gamma_Q \right].
\label{eq:DaQP}
\end{equation}
This is the analog of Eq. \ref{eq:aQHph}.  The renormalization corrections are smaller, as suggested by a $\Delta$
in their formulas.


\section{Conclusion}
\label{sec:conc}

Except in metals at low $T$,
QH theory gives the correct leading order $V$ derivatives of $F(V,T)$.  
Therefore thermal expansion and thermal shift of the bulk modulus 
are accurately described in leading order.  
QH should not be considered a theory of anharmonic effects.  In a hypothetical purely harmonic
crystal, the frequencies $\omega_Q$ are independent of $V$, but a non-zero $\gamma_Q$ does not
require explicit anharmonic forces.  The insistence that this is ``anharmonic'' only causes confusion.
The computed harmonic frequencies $\omega_Q^{\rm QH}(V_0)$ are usually close to
the measured quasiparticle frequencies $\omega_Q^{\rm QP}(V,T)$, but differ increasingly at high $T$.
However, the volume derivative needed for $\partial F/\partial V$ is dominated by the volume dependence
of the harmonic frequency $\omega_Q(V)$.  The correction from the volume derivative of 
$\omega_Q^{\rm QP}(V,T)-\omega_Q(V)$ is higher order, but not negligible at high $T$.

\section{Appendix}
\label{sec:app}

The aim is to do integrals

\begin{eqnarray}
I_{\rm el}(T)&=&-k_B\sum_K \int_0^T dT^\prime[f_K(T^\prime) \ln f_K(T^\prime) \nonumber \\
&+&(1-f_K(T^\prime))\ln(1-f_K(T^\prime))].
\label{eq:Iel}
\end{eqnarray}
\begin{eqnarray}
I_{\rm ph}(T)&=&-k_B \sum_Q \int_0^T dT^\prime [n_Q(T^\prime)\ln n_Q(T^\prime) \nonumber \\
&-&(n_Q(T^\prime)+1)\ln(n_Q(T^\prime)+1)].
\label{eq:Iph}
\end{eqnarray}
The trick is to write for electrons
\begin{equation}
T = \frac{\epsilon_K -\mu}{k_B}\frac{1}{\ln((1-f_K)/f_K)} 
\label{eq:Tel}
\end{equation}
and for phonons
\begin{equation}
T = \frac{\hbar\omega_Q}{k_B}\frac{1}{\ln((1+n_Q)/n_Q)} .
\label{eq:Tel}
\end{equation}
Then the integrals become
\begin{eqnarray}
I_{\rm el}(T)&=&-k_B\sum_K \int_{f_K(0)}^{f_K(T)} df_K [f_K \ln f_K \nonumber \\
&+&(1-f_K)\ln(1-f_K)]\frac{dT}{df_K},
\label{eq:Iel1}
\end{eqnarray}
and
\begin{eqnarray}
I_{\rm ph}(T)&=&-k_B \sum_Q \int_{n_Q(0)}^{n_Q(T)} dn_Q [n_Q\ln n_Q \nonumber \\
&-&(n_Q+1)\ln(n_Q+1)]\frac{dT}{dn_Q}.
\label{eq:Iph1}
\end{eqnarray}
Because the QH energies $\epsilon_K$ and $\omega_Q$ do not depend on $T$, the derivatives
$dT/df_K$ and $dT/dn_Q$ are easy.  The resulting formulas can be integrated by parts, and
yield the answers shown in Eqs. \ref{eq:Fbareel} and \ref{eq:Fbareph}.

\section{erratum}

The text above was already published in Modern Physics Letters B
Vol. 34, No. 2 (2020) 2050025. 
Three interesting papers (\onlinecite{Delaire2008,Bock2005,Bock2006}) should have been cited.

Stimulated by Varma {\it et al.} \cite{Varma2022}, I noticed an important error 
in this paper,  which invalidates some
of the results, but is easily corrected.  The underlying method is to find 
temperature ($T$)-dependent
corrections from electron-phonon interactions to the free energy $F(V,T)$.
The starting point is the quasiparticle (QP) energy $\epsilon_{k,{\rm QP}}=\epsilon_k(V)+
\Delta \epsilon_{k,{\rm EP}}(T)$ of electron bands.  When the electron-phonon shift
$\Delta \epsilon_{k,{\rm EP}}$ is omitted, the volume ($V$) dependence of band energies $\epsilon_k(V)$ 
gives the quasi-harmonic approximation (QHA).
The QP corrections to $\epsilon_k$ are then used to find corrections beyond QHA to properties like thermal expansion.
Another correction not given by $\Delta \epsilon_{k,{\rm EP}}$ should be included.

It is argued that the entropy $S(V,T)$ of interacting quasiparticles is given (provided quasiparticle
 energies are not too badly smeared by lifetime broadening) by using
renormalized (and generally $T$-dependent) quasiparticle energies 
 in the formulas for the entropy of noninteracting particles.

The free energy is then obtained from the entropy using 
\begin{equation}
F(V,T)=F(V,0)-\int_0^T dT^\prime S(V,T^\prime).
\label{eq:FSint}
\end{equation}
This is Eq. 25 of the main text.  
The error is failure to note that the zero-temperature piece $F(V,0)$ contains an additional correction
related to quasiparticle renormalization.   Here I argue that this correction can be easily constructed
using density-functional theory (DFT) results.
At $T=0$, $F(V,0)=E(V,0)\rightarrow E_{\rm DFT}(V) + \Delta E_{\rm EP}(V,0)$.  The electron-phonon 
correction $\Delta E_{\rm EP}$ to lowest order is found from the effective Hamiltonian
\begin{equation}
H_{\rm eff}=\sum_k \epsilon_k(V) c^\dagger_k c_k + \sum_Q \hbar\omega_Q (a^\dagger_Q a_Q +1/2) +V_{\rm ep}
\label{eq:Heff}
\end{equation}
\begin{eqnarray}
V_{\rm ep}&=& \sum_{kQ} V^{(1)}(kQ)c^\dagger_{k+Q}c_k (a_Q +a^\dagger_{-Q}) \nonumber \\
&+& \sum_{kQQ^\prime} V^{(2)}(kQQ^\prime)c^\dagger_{k+Q+Q^\prime}c_k (a_Q +a^\dagger_{-Q})
(a_{Q^\prime} +a^\dagger_{-Q^\prime}) \nonumber \\
&+& \cdots
\label{eq:Vep}
\end{eqnarray}
where $V^{(1)}$ and $V^{(2)}$ are the lowest order Taylor expansion of the DFT energy at
distorted atom coordinates $\vec{R}_\ell = \vec{\ell}+\vec{u}_\ell$.  The electron excitation
energies $\epsilon_k(V)$ for ordinary metals align well with the DFT energies $\epsilon_{k,{\rm DFT}}(V)$,
and the phonon excitations $\omega_Q$ are ordinarily given by density-functional perturbation theory (DFPT).
%
%
%
Then the perturbative correction to the ground state energy is
\begin{eqnarray}
&&\Delta E_{\rm EP}(V,0)=\sum_k \left[\langle k|V^{(2)}|k\rangle \right. \nonumber \\
&+&\left. \sum_Q
\frac{|\langle k|V^{(1)}|k+Q\rangle|^2}{\epsilon_k-\epsilon_{k+Q}} (1-f_{k+Q}) \right] f_k,
\label{eq:newcorr}
\end{eqnarray}
where $f_k$ is the Fermi distribution, here at $T=0$.
This result has the same structure as the sum of electron-phonon energy shifts $\Delta\epsilon_{k,EP}$
of occupied states $k$, given by the Allen-Heine formula \cite{Allen1976}.  But it differs because of the factor
$1-f_{k+Q}$ in the second term, which is not present in the QP energy shifts.
The additional lowest-order electron-phonon correction (eq. \ref{eq:newcorr}) to the free energy is of
similar magnitude to the corrections already discussed in the main text.  
In particular, eqs. 26, 28, 30, 32, and 37
need corrections easily derivable from eq. \ref{eq:newcorr}.



\section{acknowledgements}
I thank S. Baroni and M. L. Klein for help with the literature, and
B. Fultz, P. Ordejon, and N. K. Ravichandran for discussions.
This work was supported in part by DOE grant No. DE-FG02-08ER46550.
%


\bibliography{QH2019H}

\begin{thebibliography}{41}%
\makeatletter
\providecommand \@ifxundefined [1]{%
 \@ifx{#1\undefined}
}%
\providecommand \@ifnum [1]{%
 \ifnum #1\expandafter \@firstoftwo
 \else \expandafter \@secondoftwo
 \fi
}%
\providecommand \@ifx [1]{%
 \ifx #1\expandafter \@firstoftwo
 \else \expandafter \@secondoftwo
 \fi
}%
\providecommand \natexlab [1]{#1}%
\providecommand \enquote  [1]{``#1''}%
\providecommand \bibnamefont  [1]{#1}%
\providecommand \bibfnamefont [1]{#1}%
\providecommand \citenamefont [1]{#1}%
\providecommand \href@noop [0]{\@secondoftwo}%
\providecommand \href [0]{\begingroup \@sanitize@url \@href}%
\providecommand \@href[1]{\@@startlink{#1}\@@href}%
\providecommand \@@href[1]{\endgroup#1\@@endlink}%
\providecommand \@sanitize@url [0]{\catcode `\\12\catcode `\$12\catcode
  `\&12\catcode `\#12\catcode `\^12\catcode `\_12\catcode `\%12\relax}%
\providecommand \@@startlink[1]{}%
\providecommand \@@endlink[0]{}%
\providecommand \url  [0]{\begingroup\@sanitize@url \@url }%
\providecommand \@url [1]{\endgroup\@href {#1}{\urlprefix }}%
\providecommand \urlprefix  [0]{URL }%
\providecommand \Eprint [0]{\href }%
\providecommand \doibase [0]{http://dx.doi.org/}%
\providecommand \selectlanguage [0]{\@gobble}%
\providecommand \bibinfo  [0]{\@secondoftwo}%
\providecommand \bibfield  [0]{\@secondoftwo}%
\providecommand \translation [1]{[#1]}%
\providecommand \BibitemOpen [0]{}%
\providecommand \bibitemStop [0]{}%
\providecommand \bibitemNoStop [0]{.\EOS\space}%
\providecommand \EOS [0]{\spacefactor3000\relax}%
\providecommand \BibitemShut  [1]{\csname bibitem#1\endcsname}%
\let\auto@bib@innerbib\@empty
\bibitem [{\citenamefont {Martin}\ \emph {et~al.}(2016)\citenamefont {Martin},
  \citenamefont {Reining},\ and\ \citenamefont {Ceperley}}]{Martin2016}%
  \BibitemOpen
  \bibfield  {author} {\bibinfo {author} {\bibfnamefont {R.~M.}\ \bibnamefont
  {Martin}}, \bibinfo {author} {\bibfnamefont {L.}~\bibnamefont {Reining}}, \
  and\ \bibinfo {author} {\bibfnamefont {D.~M.}\ \bibnamefont {Ceperley}},\
  }\href {\doibase 10.1017/CBO9781139050807} {\emph {\bibinfo {title}
  {Interacting Electrons: Theory and Computational Approaches}}}\ (\bibinfo
  {publisher} {Cambridge University Press},\ \bibinfo {year}
  {2016})\BibitemShut {NoStop}%
\bibitem [{\citenamefont {Baroni}\ \emph {et~al.}(2001)\citenamefont {Baroni},
  \citenamefont {de~Gironcoli}, \citenamefont {Dal~Corso},\ and\ \citenamefont
  {Giannozzi}}]{Baroni2001}%
  \BibitemOpen
  \bibfield  {author} {\bibinfo {author} {\bibfnamefont {S.}~\bibnamefont
  {Baroni}}, \bibinfo {author} {\bibfnamefont {S.}~\bibnamefont
  {de~Gironcoli}}, \bibinfo {author} {\bibfnamefont {A.}~\bibnamefont
  {Dal~Corso}}, \ and\ \bibinfo {author} {\bibfnamefont {P.}~\bibnamefont
  {Giannozzi}},\ }\bibfield  {title} {\enquote {\bibinfo {title} {Phonons and
  related crystal properties from density-functional perturbation theory},}\
  }\href {\doibase 10.1103/RevModPhys.73.515} {\bibfield  {journal} {\bibinfo
  {journal} {Rev. Mod. Phys.}\ }\textbf {\bibinfo {volume} {73}},\ \bibinfo
  {pages} {515--562} (\bibinfo {year} {2001})}\BibitemShut {NoStop}%
\bibitem [{\citenamefont {Baroni}\ \emph {et~al.}(2010)\citenamefont {Baroni},
  \citenamefont {Giannozzi},\ and\ \citenamefont {Isaev}}]{Baroni2010}%
  \BibitemOpen
  \bibfield  {author} {\bibinfo {author} {\bibfnamefont {S.}~\bibnamefont
  {Baroni}}, \bibinfo {author} {\bibfnamefont {P.}~\bibnamefont {Giannozzi}}, \
  and\ \bibinfo {author} {\bibfnamefont {E.}~\bibnamefont {Isaev}},\ }\bibfield
   {title} {\enquote {\bibinfo {title} {Density-functional perturbation theory
  for quasi-harmonic calculations},}\ }\href@noop {} {\bibfield  {journal}
  {\bibinfo  {journal} {Rev. Mineral. Geochem.}\ }\textbf {\bibinfo {volume}
  {71}},\ \bibinfo {pages} {39} (\bibinfo {year} {2010})}\BibitemShut {NoStop}%
\bibitem [{\citenamefont {Gr{\"u}neisen}(1912)}]{Grueneisen1912}%
  \BibitemOpen
  \bibfield  {author} {\bibinfo {author} {\bibfnamefont {E.}~\bibnamefont
  {Gr{\"u}neisen}},\ }\bibfield  {title} {\enquote {\bibinfo {title} {{Theorie
  des festen Zustandes einatomiger Elemente}},}\ }\href {\doibase
  10.1002/andp.19123441202} {\bibfield  {journal} {\bibinfo  {journal} {Annalen
  der Physik}\ }\textbf {\bibinfo {volume} {344}},\ \bibinfo {pages} {257--306}
  (\bibinfo {year} {1912})}\BibitemShut {NoStop}%
\bibitem [{\citenamefont {Gr{\"u}neisen}(1926)}]{Grueneisen1926}%
  \BibitemOpen
  \bibfield  {author} {\bibinfo {author} {\bibfnamefont {E.}~\bibnamefont
  {Gr{\"u}neisen}},\ }\bibfield  {title} {\enquote {\bibinfo {title} {Zustand
  des festen {K\"orpers}},}\ }in\ \href@noop {} {\emph {\bibinfo {booktitle}
  {Handbuch der Physik}}},\ \bibinfo {editor} {edited by\ \bibinfo {editor}
  {\bibfnamefont {H.}~\bibnamefont {Geiger}}\ and\ \bibinfo {editor}
  {\bibfnamefont {K.}~\bibnamefont {Scheel}}}\ (\bibinfo  {publisher}
  {Springer},\ \bibinfo {address} {Berlin},\ \bibinfo {year} {1926})\ pp.\
  \bibinfo {pages} {1--52},\ \bibinfo {note} {{A}n {E}nglish translation is
  available at http://www.dtic.mil/dtic/tr/fulltext/u2/215056.pdf}\BibitemShut
  {NoStop}%
\bibitem [{\citenamefont {{Otero-de-la-Roza}}\ \emph
  {et~al.}(2011)\citenamefont {{Otero-de-la-Roza}}, \citenamefont
  {Abbasi-P\'erez},\ and\ \citenamefont {Lua\~na}}]{Otero2011}%
  \BibitemOpen
  \bibfield  {author} {\bibinfo {author} {\bibfnamefont {A.}~\bibnamefont
  {{Otero-de-la-Roza}}}, \bibinfo {author} {\bibfnamefont {D.}~\bibnamefont
  {Abbasi-P\'erez}}, \ and\ \bibinfo {author} {\bibfnamefont {V.}~\bibnamefont
  {Lua\~na}},\ }\bibfield  {title} {\enquote {\bibinfo {title} {{GIBBS2: A new
  version of the quasiharmonic model code. II. Models for solid-state
  thermodynamics, features and implementation}},}\ }\href@noop {} {\bibfield
  {journal} {\bibinfo  {journal} {Computer Physics Communications}\ }\textbf
  {\bibinfo {volume} {182}},\ \bibinfo {pages} {2232 -- 2248} (\bibinfo {year}
  {2011})}\BibitemShut {NoStop}%
\bibitem [{\citenamefont {{Otero-de-la-Roza}}\ and\ \citenamefont
  {Lua\~na}(2011{\natexlab{a}})}]{Otero-a-2011}%
  \BibitemOpen
  \bibfield  {author} {\bibinfo {author} {\bibfnamefont {A.}~\bibnamefont
  {{Otero-de-la-Roza}}}\ and\ \bibinfo {author} {\bibfnamefont
  {V.}~\bibnamefont {Lua\~na}},\ }\bibfield  {title} {\enquote {\bibinfo
  {title} {{Treatment of first-principles data for predictive quasiharmonic
  thermodynamics of solids: The case of MgO}},}\ }\href {\doibase
  10.1103/PhysRevB.84.024109} {\bibfield  {journal} {\bibinfo  {journal} {Phys.
  Rev. B}\ }\textbf {\bibinfo {volume} {84}},\ \bibinfo {pages} {024109}
  (\bibinfo {year} {2011}{\natexlab{a}})}\BibitemShut {NoStop}%
\bibitem [{\citenamefont {{Otero-de-la-Roza}}\ and\ \citenamefont
  {Lua\~na}(2011{\natexlab{b}})}]{Otero-b-2011}%
  \BibitemOpen
  \bibfield  {author} {\bibinfo {author} {\bibfnamefont {A.}~\bibnamefont
  {{Otero-de-la-Roza}}}\ and\ \bibinfo {author} {\bibfnamefont {V\'{\i}ctor}\
  \bibnamefont {Lua\~na}},\ }\bibfield  {title} {\enquote {\bibinfo {title}
  {Equations of state and thermodynamics of solids using empirical corrections
  in the quasiharmonic approximation},}\ }\href {\doibase
  10.1103/PhysRevB.84.184103} {\bibfield  {journal} {\bibinfo  {journal} {Phys.
  Rev. B}\ }\textbf {\bibinfo {volume} {84}},\ \bibinfo {pages} {184103}
  (\bibinfo {year} {2011}{\natexlab{b}})}\BibitemShut {NoStop}%
\bibitem [{\citenamefont {Mikura}(1941)}]{Mikura1941}%
  \BibitemOpen
  \bibfield  {author} {\bibinfo {author} {\bibfnamefont {Z.}~\bibnamefont
  {Mikura}},\ }\bibfield  {title} {\enquote {\bibinfo {title} {Contribution of
  the conduction electrons in a metal to the thermal expansion},}\ }\href@noop
  {} {\bibfield  {journal} {\bibinfo  {journal} {Proc. Phys. Math. Soc. Japan}\
  }\textbf {\bibinfo {volume} {23}},\ \bibinfo {pages} {309} (\bibinfo {year}
  {1941})}\BibitemShut {NoStop}%
\bibitem [{\citenamefont {Visvanathan}(1951)}]{Visvanathan1951}%
  \BibitemOpen
  \bibfield  {author} {\bibinfo {author} {\bibfnamefont {S.}~\bibnamefont
  {Visvanathan}},\ }\bibfield  {title} {\enquote {\bibinfo {title} {Thermal
  expansion at low temperatures},}\ }\href {\doibase 10.1103/PhysRev.81.626.3}
  {\bibfield  {journal} {\bibinfo  {journal} {Phys. Rev.}\ }\textbf {\bibinfo
  {volume} {81}},\ \bibinfo {pages} {626--627} (\bibinfo {year}
  {1951})}\BibitemShut {NoStop}%
\bibitem [{\citenamefont {Erba}\ \emph {et~al.}(2015)\citenamefont {Erba},
  \citenamefont {Shahrokhi}, \citenamefont {Moradian},\ and\ \citenamefont
  {Dovesi}}]{Erba2015}%
  \BibitemOpen
  \bibfield  {author} {\bibinfo {author} {\bibfnamefont {A.}~\bibnamefont
  {Erba}}, \bibinfo {author} {\bibfnamefont {M.}~\bibnamefont {Shahrokhi}},
  \bibinfo {author} {\bibfnamefont {R.}~\bibnamefont {Moradian}}, \ and\
  \bibinfo {author} {\bibfnamefont {R.}~\bibnamefont {Dovesi}},\ }\bibfield
  {title} {\enquote {\bibinfo {title} {On how differently the quasi-harmonic
  approximation works for two isostructural crystals: Thermal properties of
  periclase and lime},}\ }\href@noop {} {\bibfield  {journal} {\bibinfo
  {journal} {J. Chem. Phys.}\ }\textbf {\bibinfo {volume} {142}},\ \bibinfo
  {pages} {044114} (\bibinfo {year} {2015})}\BibitemShut {NoStop}%
\bibitem [{\citenamefont {Kim}\ \emph {et~al.}(2018)\citenamefont {Kim},
  \citenamefont {Hellman}, \citenamefont {Herriman}, \citenamefont {Smith},
  \citenamefont {Lin}, \citenamefont {Shulumba}, \citenamefont {Niedziela},
  \citenamefont {Li}, \citenamefont {Abernathy},\ and\ \citenamefont
  {Fultz}}]{Kim2018}%
  \BibitemOpen
  \bibfield  {author} {\bibinfo {author} {\bibfnamefont {D.~S.}\ \bibnamefont
  {Kim}}, \bibinfo {author} {\bibfnamefont {O.}~\bibnamefont {Hellman}},
  \bibinfo {author} {\bibfnamefont {J.}~\bibnamefont {Herriman}}, \bibinfo
  {author} {\bibfnamefont {H.~L.}\ \bibnamefont {Smith}}, \bibinfo {author}
  {\bibfnamefont {J.~Y.~Y.}\ \bibnamefont {Lin}}, \bibinfo {author}
  {\bibfnamefont {N.}~\bibnamefont {Shulumba}}, \bibinfo {author}
  {\bibfnamefont {J.~L.}\ \bibnamefont {Niedziela}}, \bibinfo {author}
  {\bibfnamefont {C.~W.}\ \bibnamefont {Li}}, \bibinfo {author} {\bibfnamefont
  {D.~L.}\ \bibnamefont {Abernathy}}, \ and\ \bibinfo {author} {\bibfnamefont
  {B.}~\bibnamefont {Fultz}},\ }\bibfield  {title} {\enquote {\bibinfo {title}
  {Nuclear quantum effect with pure anharmonicity and the anomalous thermal
  expansion of silicon},}\ }\href {\doibase 10.1073/pnas.1707745115} {\bibfield
   {journal} {\bibinfo  {journal} {Proc. Nat. Acad. Sci.}\ }\textbf {\bibinfo
  {volume} {115}},\ \bibinfo {pages} {1992--1997} (\bibinfo {year}
  {2018})}\BibitemShut {NoStop}%
\bibitem [{\citenamefont {Shen}\ \emph {et~al.}(2020)\citenamefont {Shen},
  \citenamefont {Saunders}, \citenamefont {Bernal}, \citenamefont {Abernathy},
  \citenamefont {Manley},\ and\ \citenamefont {Fultz}}]{Fultz2019}%
  \BibitemOpen
  \bibfield  {author} {\bibinfo {author} {\bibfnamefont {Y.}~\bibnamefont
  {Shen}}, \bibinfo {author} {\bibfnamefont {C.~N.}\ \bibnamefont {Saunders}},
  \bibinfo {author} {\bibfnamefont {C.~M.}\ \bibnamefont {Bernal}}, \bibinfo
  {author} {\bibfnamefont {D.~L.}\ \bibnamefont {Abernathy}}, \bibinfo {author}
  {\bibfnamefont {M.~E.}\ \bibnamefont {Manley}}, \ and\ \bibinfo {author}
  {\bibfnamefont {B.}~\bibnamefont {Fultz}},\ }\bibfield  {title} {\enquote
  {\bibinfo {title} {Anharmonic origin of the giant thermal expansion of
  nabr},}\ }\href {\doibase 10.1103/PhysRevLett.125.085504} {\bibfield
  {journal} {\bibinfo  {journal} {Phys. Rev. Lett.}\ }\textbf {\bibinfo
  {volume} {125}},\ \bibinfo {pages} {085504} (\bibinfo {year}
  {2020})}\BibitemShut {NoStop}%
\bibitem [{\citenamefont {Delaire}\ \emph {et~al.}(2011)\citenamefont
  {Delaire}, \citenamefont {Ma}, \citenamefont {Marty}, \citenamefont {May},
  \citenamefont {McGuire}, \citenamefont {Du}, \citenamefont {Singh},
  \citenamefont {Podlesnyak}, \citenamefont {Ehlers}, \citenamefont {Lumsden},\
  and\ \citenamefont {Sales}}]{Delaire2011}%
  \BibitemOpen
  \bibfield  {author} {\bibinfo {author} {\bibfnamefont {O.}~\bibnamefont
  {Delaire}}, \bibinfo {author} {\bibfnamefont {J.}~\bibnamefont {Ma}},
  \bibinfo {author} {\bibfnamefont {K.}~\bibnamefont {Marty}}, \bibinfo
  {author} {\bibfnamefont {A.~F.}\ \bibnamefont {May}}, \bibinfo {author}
  {\bibfnamefont {M.~A.}\ \bibnamefont {McGuire}}, \bibinfo {author}
  {\bibfnamefont {M-H.}\ \bibnamefont {Du}}, \bibinfo {author} {\bibfnamefont
  {D.~J.}\ \bibnamefont {Singh}}, \bibinfo {author} {\bibfnamefont
  {A.}~\bibnamefont {Podlesnyak}}, \bibinfo {author} {\bibfnamefont
  {G.}~\bibnamefont {Ehlers}}, \bibinfo {author} {\bibfnamefont {M.~D.}\
  \bibnamefont {Lumsden}}, \ and\ \bibinfo {author} {\bibfnamefont {B.~C.}\
  \bibnamefont {Sales}},\ }\bibfield  {title} {\enquote {\bibinfo {title}
  {{Giant anharmonic phonon scattering in PbTe}},}\ }\href
  {http://proxy.library.stonybrook.edu/login?url=http://search.ebscohost.com/login.aspx?direct=true&db=a9h&AN=63004994&site=ehost-live&scope=site}
  {\bibfield  {journal} {\bibinfo  {journal} {Nature Materials}\ }\textbf
  {\bibinfo {volume} {10}},\ \bibinfo {pages} {614 -- 619} (\bibinfo {year}
  {2011})}\BibitemShut {NoStop}%
\bibitem [{\citenamefont {Skelton}\ \emph {et~al.}(2014)\citenamefont
  {Skelton}, \citenamefont {Parker}, \citenamefont {Togo}, \citenamefont
  {Tanaka},\ and\ \citenamefont {Walsh}}]{Skelton2014}%
  \BibitemOpen
  \bibfield  {author} {\bibinfo {author} {\bibfnamefont {J.~M.}\ \bibnamefont
  {Skelton}}, \bibinfo {author} {\bibfnamefont {S.~C.}\ \bibnamefont {Parker}},
  \bibinfo {author} {\bibfnamefont {A.}~\bibnamefont {Togo}}, \bibinfo {author}
  {\bibfnamefont {I.}~\bibnamefont {Tanaka}}, \ and\ \bibinfo {author}
  {\bibfnamefont {A.}~\bibnamefont {Walsh}},\ }\bibfield  {title} {\enquote
  {\bibinfo {title} {{Thermal physics of the lead chalcogenides PbS, PbSe, and
  PbTe from first principles}},}\ }\href {\doibase 10.1103/PhysRevB.89.205203}
  {\bibfield  {journal} {\bibinfo  {journal} {Phys. Rev. B}\ }\textbf {\bibinfo
  {volume} {89}},\ \bibinfo {pages} {205203} (\bibinfo {year}
  {2014})}\BibitemShut {NoStop}%
\bibitem [{\citenamefont {Houston}\ \emph {et~al.}(1968)\citenamefont
  {Houston}, \citenamefont {Strakna},\ and\ \citenamefont
  {Belson}}]{Houston1968}%
  \BibitemOpen
  \bibfield  {author} {\bibinfo {author} {\bibfnamefont {B.}~\bibnamefont
  {Houston}}, \bibinfo {author} {\bibfnamefont {R.~E.}\ \bibnamefont
  {Strakna}}, \ and\ \bibinfo {author} {\bibfnamefont {H.~S.}\ \bibnamefont
  {Belson}},\ }\bibfield  {title} {\enquote {\bibinfo {title} {{Elastic
  constants, thermal expansion, and Debye temperature of lead telluride}},}\
  }\href {\doibase 10.1063/1.1656874} {\bibfield  {journal} {\bibinfo
  {journal} {J. Appl. Phys.}\ }\textbf {\bibinfo {volume} {39}},\ \bibinfo
  {pages} {3913--3916} (\bibinfo {year} {1968})}\BibitemShut {NoStop}%
\bibitem [{\citenamefont {Barron}(1961)}]{Barron1961}%
  \BibitemOpen
  \bibfield  {author} {\bibinfo {author} {\bibfnamefont {T.~H.~K.}\
  \bibnamefont {Barron}},\ }\bibfield  {title} {\enquote {\bibinfo {title}
  {Equation of state and thermodynamic properties},}\ }in\ \href@noop {} {\emph
  {\bibinfo {booktitle} {Proceedings of the VIIth International Conference on
  Low Temperature Physics}}},\ \bibinfo {editor} {edited by\ \bibinfo {editor}
  {\bibfnamefont {G.~M.}\ \bibnamefont {Graham}}\ and\ \bibinfo {editor}
  {\bibfnamefont {A.~C.~Hollis}\ \bibnamefont {Hallett}}}\ (\bibinfo
  {publisher} {U. of Toronto Press},\ \bibinfo {address} {Toronto},\ \bibinfo
  {year} {1961})\ pp.\ \bibinfo {pages} {655 -- 670}\BibitemShut {NoStop}%
\bibitem [{\citenamefont {Leibfried}\ and\ \citenamefont
  {Ludwig}(1961)}]{Leibfried1961}%
  \BibitemOpen
  \bibfield  {author} {\bibinfo {author} {\bibfnamefont {G.}~\bibnamefont
  {Leibfried}}\ and\ \bibinfo {author} {\bibfnamefont {W.}~\bibnamefont
  {Ludwig}},\ }\bibfield  {title} {\enquote {\bibinfo {title} {Theory of
  anharmonic effects in crystals},}\ }in\ \href@noop {} {\emph {\bibinfo
  {booktitle} {Solid State Physics}}},\ Vol.~\bibinfo {volume} {12},\ \bibinfo
  {editor} {edited by\ \bibinfo {editor} {\bibfnamefont {F.}~\bibnamefont
  {Seitz}}\ and\ \bibinfo {editor} {\bibfnamefont {D.}~\bibnamefont
  {Turnbull}}}\ (\bibinfo  {publisher} {Academic Press},\ \bibinfo {address}
  {New York},\ \bibinfo {year} {1961})\ pp.\ \bibinfo {pages} {276 --
  444}\BibitemShut {NoStop}%
\bibitem [{\citenamefont {Cowley}\ \emph {et~al.}(1966)\citenamefont {Cowley},
  \citenamefont {Cowley},\ and\ \citenamefont {W.}}]{Cowley1966I}%
  \BibitemOpen
  \bibfield  {author} {\bibinfo {author} {\bibfnamefont {E.~R.}\ \bibnamefont
  {Cowley}}, \bibinfo {author} {\bibfnamefont {R.~A.}\ \bibnamefont {Cowley}},
  \ and\ \bibinfo {author} {\bibfnamefont {Cochran}\ \bibnamefont {W.}},\
  }\bibfield  {title} {\enquote {\bibinfo {title} {Anharmonic interactions in
  alkali halides {I}},}\ }\href@noop {} {\bibfield  {journal} {\bibinfo
  {journal} {Proc. R. Soc. Lond. A}\ }\textbf {\bibinfo {volume} {287}},\
  \bibinfo {pages} {259--280} (\bibinfo {year} {1966})}\BibitemShut {NoStop}%
\bibitem [{\citenamefont {Cowley}(1968)}]{Cowley1968}%
  \BibitemOpen
  \bibfield  {author} {\bibinfo {author} {\bibfnamefont {R.~A.}\ \bibnamefont
  {Cowley}},\ }\bibfield  {title} {\enquote {\bibinfo {title} {Anharmonic
  crystals},}\ }\href {http://stacks.iop.org/0034-4885/31/i=1/a=303} {\bibfield
   {journal} {\bibinfo  {journal} {Rep. Prog. Phys.}\ }\textbf {\bibinfo
  {volume} {31}},\ \bibinfo {pages} {123} (\bibinfo {year} {1968})}\BibitemShut
  {NoStop}%
\bibitem [{\citenamefont {Wallace}(1972)}]{Wallace1972}%
  \BibitemOpen
  \bibfield  {author} {\bibinfo {author} {\bibfnamefont {D.~C.}\ \bibnamefont
  {Wallace}},\ }\href@noop {} {\emph {\bibinfo {title} {Thermodynamics of
  Crystals}}}\ (\bibinfo  {publisher} {Wiley},\ \bibinfo {address} {New York},\
  \bibinfo {year} {1972})\ \bibinfo {note} {, reprinted by Dover Publications,
  1998}\BibitemShut {NoStop}%
\bibitem [{\citenamefont {Allen}\ and\ \citenamefont
  {Heine}(1976)}]{Allen1976}%
  \BibitemOpen
  \bibfield  {author} {\bibinfo {author} {\bibfnamefont {P.~B.}\ \bibnamefont
  {Allen}}\ and\ \bibinfo {author} {\bibfnamefont {V.}~\bibnamefont {Heine}},\
  }\bibfield  {title} {\enquote {\bibinfo {title} {Theory of the temperature
  dependence of electronic band structures},}\ }\href {\doibase
  10.1088/0022-3719/9/12/013} {\bibfield  {journal} {\bibinfo  {journal} {J.
  Phys. C: Solid State Physics}\ }\textbf {\bibinfo {volume} {9}},\ \bibinfo
  {pages} {2305--2312} (\bibinfo {year} {1976})}\BibitemShut {NoStop}%
\bibitem [{\citenamefont {Ponc\'e}\ \emph {et~al.}(2015)\citenamefont
  {Ponc\'e}, \citenamefont {Gillet}, \citenamefont {Laflamme~Janssen},
  \citenamefont {Marini}, \citenamefont {Verstraete},\ and\ \citenamefont
  {Gonze}}]{Ponce2015}%
  \BibitemOpen
  \bibfield  {author} {\bibinfo {author} {\bibfnamefont {S.}~\bibnamefont
  {Ponc\'e}}, \bibinfo {author} {\bibfnamefont {Y.}~\bibnamefont {Gillet}},
  \bibinfo {author} {\bibfnamefont {J.}~\bibnamefont {Laflamme~Janssen}},
  \bibinfo {author} {\bibfnamefont {A.}~\bibnamefont {Marini}}, \bibinfo
  {author} {\bibfnamefont {M.}~\bibnamefont {Verstraete}}, \ and\ \bibinfo
  {author} {\bibfnamefont {X.}~\bibnamefont {Gonze}},\ }\bibfield  {title}
  {\enquote {\bibinfo {title} {Temperature dependence of the electronic
  structure of semiconductors and insulators},}\ }\href@noop {} {\bibfield
  {journal} {\bibinfo  {journal} {J. Chem. Phys.}\ }\textbf {\bibinfo {volume}
  {143}},\ \bibinfo {pages} {102813} (\bibinfo {year} {2015})}\BibitemShut
  {NoStop}%
\bibitem [{\citenamefont {Grimvall}(1981)}]{Grimvall1981}%
  \BibitemOpen
  \bibfield  {author} {\bibinfo {author} {\bibfnamefont {G.}~\bibnamefont
  {Grimvall}},\ }\href@noop {} {\emph {\bibinfo {title} {The Electron-Phonon
  Interaction in Metals (Selected Topics in Solid State Physics XVI)}}}\
  (\bibinfo  {publisher} {North Holland},\ \bibinfo {address} {Amsterdam},\
  \bibinfo {year} {1981})\BibitemShut {NoStop}%
\bibitem [{\citenamefont {Grimvall}(1986)}]{Grimvall1986}%
  \BibitemOpen
  \bibfield  {author} {\bibinfo {author} {\bibfnamefont {G.}~\bibnamefont
  {Grimvall}},\ }\href@noop {} {\emph {\bibinfo {title} {Thermophysical
  Properties of Materials}}}\ (\bibinfo  {publisher} {North Holland},\ \bibinfo
  {address} {Amsterdam},\ \bibinfo {year} {1986})\BibitemShut {NoStop}%
\bibitem [{\citenamefont {Stern}(1958)}]{Stern1958}%
  \BibitemOpen
  \bibfield  {author} {\bibinfo {author} {\bibfnamefont {E.~A.}\ \bibnamefont
  {Stern}},\ }\bibfield  {title} {\enquote {\bibinfo {title} {Theory of the
  anharmonic properties of solids},}\ }\href {\doibase 10.1103/PhysRev.111.786}
  {\bibfield  {journal} {\bibinfo  {journal} {Phys. Rev.}\ }\textbf {\bibinfo
  {volume} {111}},\ \bibinfo {pages} {786--797} (\bibinfo {year}
  {1958})}\BibitemShut {NoStop}%
\bibitem [{\citenamefont {Ziman}(1960)}]{Ziman1960}%
  \BibitemOpen
  \bibfield  {author} {\bibinfo {author} {\bibfnamefont {J.}~\bibnamefont
  {Ziman}},\ }\href@noop {} {\emph {\bibinfo {title} {Electrons and Phonons}}}\
  (\bibinfo  {publisher} {Oxford University Press},\ \bibinfo {address} {New
  York},\ \bibinfo {year} {1960})\BibitemShut {NoStop}%
\bibitem [{\citenamefont {Pavone}\ and\ \citenamefont
  {Baroni}(1994)}]{Pavone1994}%
  \BibitemOpen
  \bibfield  {author} {\bibinfo {author} {\bibfnamefont {P.}~\bibnamefont
  {Pavone}}\ and\ \bibinfo {author} {\bibfnamefont {S.}~\bibnamefont
  {Baroni}},\ }\bibfield  {title} {\enquote {\bibinfo {title} {Dependence of
  the crystal lattice constant on isotopic composition: Theory and ab initio
  calculations for {C, Si, and Ge}},}\ }\href {\doibase
  https://doi.org/10.1016/0038-1098(94)90154-6} {\bibfield  {journal} {\bibinfo
   {journal} {Solid State Commun.}\ }\textbf {\bibinfo {volume} {90}},\
  \bibinfo {pages} {295 -- 297} (\bibinfo {year} {1994})}\BibitemShut {NoStop}%
\bibitem [{\citenamefont {Huang}\ \emph {et~al.}(2016)\citenamefont {Huang},
  \citenamefont {Lu}, \citenamefont {Tennessen},\ and\ \citenamefont
  {Rondinelli}}]{Huang2016}%
  \BibitemOpen
  \bibfield  {author} {\bibinfo {author} {\bibfnamefont {Liang-Feng}\
  \bibnamefont {Huang}}, \bibinfo {author} {\bibfnamefont {Xue-Zeng}\
  \bibnamefont {Lu}}, \bibinfo {author} {\bibfnamefont {E.}~\bibnamefont
  {Tennessen}}, \ and\ \bibinfo {author} {\bibfnamefont {J.~M.}\ \bibnamefont
  {Rondinelli}},\ }\bibfield  {title} {\enquote {\bibinfo {title} {An efficient
  ab-initio quasiharmonic approach for the thermodynamics of solids},}\ }\href
  {\doibase https://doi.org/10.1016/j.commatsci.2016.04.012} {\bibfield
  {journal} {\bibinfo  {journal} {Computational Materials Science}\ }\textbf
  {\bibinfo {volume} {120}},\ \bibinfo {pages} {84 -- 93} (\bibinfo {year}
  {2016})}\BibitemShut {NoStop}%
\bibitem [{\citenamefont {Wallace}(2002)}]{Wallace2002}%
  \BibitemOpen
  \bibfield  {author} {\bibinfo {author} {\bibfnamefont {D.~C.}\ \bibnamefont
  {Wallace}},\ }\href@noop {} {\emph {\bibinfo {title} {Statistical Physics of
  Crystals and Liquids}}}\ (\bibinfo  {publisher} {World Scientific},\ \bibinfo
  {address} {Singapore},\ \bibinfo {year} {2002})\BibitemShut {NoStop}%
\bibitem [{\citenamefont {Wu}\ and\ \citenamefont
  {Wentzcovitch}(2011)}]{Wu2011}%
  \BibitemOpen
  \bibfield  {author} {\bibinfo {author} {\bibfnamefont {Zhongqing}\
  \bibnamefont {Wu}}\ and\ \bibinfo {author} {\bibfnamefont {Renata~M.}\
  \bibnamefont {Wentzcovitch}},\ }\bibfield  {title} {\enquote {\bibinfo
  {title} {Quasiharmonic thermal elasticity of crystals: An analytical
  approach},}\ }\href {\doibase 10.1103/PhysRevB.83.184115} {\bibfield
  {journal} {\bibinfo  {journal} {Phys. Rev. B}\ }\textbf {\bibinfo {volume}
  {83}},\ \bibinfo {pages} {184115} (\bibinfo {year} {2011})}\BibitemShut
  {NoStop}%
\bibitem [{\citenamefont {Davies}(1974)}]{Davies1974}%
  \BibitemOpen
  \bibfield  {author} {\bibinfo {author} {\bibfnamefont {G.~F.}\ \bibnamefont
  {Davies}},\ }\bibfield  {title} {\enquote {\bibinfo {title} {{Effective
  elastic moduli under hydrostatic stress - I. quasi-harmonic theory}},}\
  }\href@noop {} {\bibfield  {journal} {\bibinfo  {journal} {J. Phys. Chem.
  Solids}\ }\textbf {\bibinfo {volume} {35}},\ \bibinfo {pages} {1513 -- 1520}
  (\bibinfo {year} {1974})}\BibitemShut {NoStop}%
\bibitem [{\citenamefont {Kantorovich}(1995)}]{Kantorovich1995-II}%
  \BibitemOpen
  \bibfield  {author} {\bibinfo {author} {\bibfnamefont {L.~N.}\ \bibnamefont
  {Kantorovich}},\ }\bibfield  {title} {\enquote {\bibinfo {title}
  {{Thermoelastic properties of perfect crystals with nonprimitive lattices.
  II. Application to KCl and NaCl}},}\ }\href {\doibase
  10.1103/PhysRevB.51.3520} {\bibfield  {journal} {\bibinfo  {journal} {Phys.
  Rev. B}\ }\textbf {\bibinfo {volume} {51}},\ \bibinfo {pages} {3535--3548}
  (\bibinfo {year} {1995})}\BibitemShut {NoStop}%
\bibitem [{\citenamefont {Werthamer}(1970)}]{Werthamer1970}%
  \BibitemOpen
  \bibfield  {author} {\bibinfo {author} {\bibfnamefont {N.~R.}\ \bibnamefont
  {Werthamer}},\ }\bibfield  {title} {\enquote {\bibinfo {title}
  {Self-consistent phonon formulation of anharmonic lattice dynamics},}\ }\href
  {\doibase 10.1103/PhysRevB.1.572} {\bibfield  {journal} {\bibinfo  {journal}
  {Phys. Rev. B}\ }\textbf {\bibinfo {volume} {1}},\ \bibinfo {pages}
  {572--581} (\bibinfo {year} {1970})}\BibitemShut {NoStop}%
\bibitem [{\citenamefont {G{\"o}tze}\ and\ \citenamefont
  {Michel}(1968)}]{Gotze1968}%
  \BibitemOpen
  \bibfield  {author} {\bibinfo {author} {\bibfnamefont {W.}~\bibnamefont
  {G{\"o}tze}}\ and\ \bibinfo {author} {\bibfnamefont {K.~H.}\ \bibnamefont
  {Michel}},\ }\bibfield  {title} {\enquote {\bibinfo {title} {Elastic
  constants of nonionic anharmonic crystals},}\ }\href@noop {} {\bibfield
  {journal} {\bibinfo  {journal} {Z. Phys. B}\ }\textbf {\bibinfo {volume}
  {217}},\ \bibinfo {pages} {170--187} (\bibinfo {year} {1968})}\BibitemShut
  {NoStop}%
\bibitem [{\citenamefont {Lifshitz}\ and\ \citenamefont
  {Pitaevsky}(1980)}]{Lifshitz1980}%
  \BibitemOpen
  \bibfield  {author} {\bibinfo {author} {\bibfnamefont {E.~M.}\ \bibnamefont
  {Lifshitz}}\ and\ \bibinfo {author} {\bibfnamefont {L.~P.}\ \bibnamefont
  {Pitaevsky}},\ }\href@noop {} {\emph {\bibinfo {title} {Statistical Physics,
  3rd Edition}}}\ (\bibinfo  {publisher} {Pergamon Press},\ \bibinfo {address}
  {Oxford},\ \bibinfo {year} {1980})\BibitemShut {NoStop}%
\bibitem [{\citenamefont {Barron}(1963)}]{Barron1963}%
  \BibitemOpen
  \bibfield  {author} {\bibinfo {author} {\bibfnamefont {T.~H.~K.}\
  \bibnamefont {Barron}},\ }\bibfield  {title} {\enquote {\bibinfo {title}
  {Thermodynamic properties and effective vibrational spectra of an anharmonic
  crystal},}\ }in\ \href@noop {} {\emph {\bibinfo {booktitle} {Lattice
  dynamics: Proceedings of the international conference held at
  {Copenhagen}}}},\ \bibinfo {editor} {edited by\ \bibinfo {editor}
  {\bibfnamefont {R.~F.}\ \bibnamefont {Wallis}}}\ (\bibinfo  {publisher}
  {Pergamon Press},\ \bibinfo {address} {Oxford},\ \bibinfo {year} {1963})\
  pp.\ \bibinfo {pages} {247 -- 254}\BibitemShut {NoStop}%
\bibitem [{\citenamefont {Delaire}\ \emph {et~al.}(2008)\citenamefont
  {Delaire}, \citenamefont {Lucas}, \citenamefont {Mu\~noz}, \citenamefont
  {Kresch},\ and\ \citenamefont {Fultz}}]{Delaire2008}%
  \BibitemOpen
  \bibfield  {author} {\bibinfo {author} {\bibfnamefont {O.}~\bibnamefont
  {Delaire}}, \bibinfo {author} {\bibfnamefont {M.~S.}\ \bibnamefont {Lucas}},
  \bibinfo {author} {\bibfnamefont {J.~A.}\ \bibnamefont {Mu\~noz}}, \bibinfo
  {author} {\bibfnamefont {M.}~\bibnamefont {Kresch}}, \ and\ \bibinfo {author}
  {\bibfnamefont {B.}~\bibnamefont {Fultz}},\ }\bibfield  {title} {\enquote
  {\bibinfo {title} {Adiabatic electron-phonon interaction and high-temperature
  thermodynamics of {A15} compounds},}\ }\href@noop {} {\bibfield  {journal}
  {\bibinfo  {journal} {Phys. Rev. Lett.}\ }\textbf {\bibinfo {volume} {101}},\
  \bibinfo {pages} {105504} (\bibinfo {year} {2008})}\BibitemShut {NoStop}%
\bibitem [{\citenamefont {Bock}\ \emph {et~al.}(2005)\citenamefont {Bock},
  \citenamefont {Coffey},\ and\ \citenamefont {C.}}]{Bock2005}%
  \BibitemOpen
  \bibfield  {author} {\bibinfo {author} {\bibfnamefont {N.}~\bibnamefont
  {Bock}}, \bibinfo {author} {\bibfnamefont {D.}~\bibnamefont {Coffey}}, \ and\
  \bibinfo {author} {\bibfnamefont {Wallace~D.}\ \bibnamefont {C.}},\
  }\bibfield  {title} {\enquote {\bibinfo {title} {Nonadiabatic contributions
  to the free energy from the electron-phonon interaction in {Na, K, Al, and
  Pb}},}\ }\href@noop {} {\bibfield  {journal} {\bibinfo  {journal} {Phys. Rev.
  B}\ }\textbf {\bibinfo {volume} {72}},\ \bibinfo {pages} {155120} (\bibinfo
  {year} {2005})}\BibitemShut {NoStop}%
\bibitem [{\citenamefont {Bock}\ \emph {et~al.}(2006)\citenamefont {Bock},
  \citenamefont {C.},\ and\ \citenamefont {Coffey}}]{Bock2006}%
  \BibitemOpen
  \bibfield  {author} {\bibinfo {author} {\bibfnamefont {N.}~\bibnamefont
  {Bock}}, \bibinfo {author} {\bibfnamefont {Wallace~D.}\ \bibnamefont {C.}}, \
  and\ \bibinfo {author} {\bibfnamefont {D.}~\bibnamefont {Coffey}},\
  }\bibfield  {title} {\enquote {\bibinfo {title} {Adiabatic and nonadiabatic
  contributions to the free energy from the electron-phonon interaction for
  {Na, K, Al, and Pb}},}\ }\href@noop {} {\bibfield  {journal} {\bibinfo
  {journal} {Phys. Rev. B}\ }\textbf {\bibinfo {volume} {73}},\ \bibinfo
  {pages} {055114} (\bibinfo {year} {2006})}\BibitemShut {NoStop}%
\bibitem [{\citenamefont {Varma}\ \emph {et~al.}()\citenamefont {Varma},
  \citenamefont {Paul}, \citenamefont {Itale}, \citenamefont {Pable},
  \citenamefont {Tibrewala}, \citenamefont {Dodal}, \citenamefont {Yerunkar},
  \citenamefont {Bhaumik}, \citenamefont {Shah}, \citenamefont {Gururajan},\
  and\ \citenamefont {Prasanna}}]{Varma2022}%
  \BibitemOpen
  \bibfield  {author} {\bibinfo {author} {\bibfnamefont {A.~R.}\ \bibnamefont
  {Varma}}, \bibinfo {author} {\bibfnamefont {S.}~\bibnamefont {Paul}},
  \bibinfo {author} {\bibfnamefont {A.}~\bibnamefont {Itale}}, \bibinfo
  {author} {\bibfnamefont {P.}~\bibnamefont {Pable}}, \bibinfo {author}
  {\bibfnamefont {R.}~\bibnamefont {Tibrewala}}, \bibinfo {author}
  {\bibfnamefont {S.}~\bibnamefont {Dodal}}, \bibinfo {author} {\bibfnamefont
  {H.}~\bibnamefont {Yerunkar}}, \bibinfo {author} {\bibfnamefont
  {S.}~\bibnamefont {Bhaumik}}, \bibinfo {author} {\bibfnamefont
  {V.}~\bibnamefont {Shah}}, \bibinfo {author} {\bibfnamefont {M.~P.}\
  \bibnamefont {Gururajan}}, \ and\ \bibinfo {author} {\bibfnamefont
  {T.~R.~S.}\ \bibnamefont {Prasanna}},\ }\bibfield  {title} {\enquote
  {\bibinfo {title} {Electron-phonon interaction contribution to the total
  energy of group {IV} semiconductor polymorphs: Evaluation and
  implications},}\ }\href@noop {} {\ }\bibinfo {note}
  {{a}r{X}iv:2204.08321}\BibitemShut {NoStop}%
\end{thebibliography}%

\end{document}